\documentclass [aps,prd,reprint,eqsecnum,nofootinbib]  {revtex4-1}
\usepackage{amssymb}
\usepackage{amsmath}

\def\SO{\sf{SO}}
\def\OSp{\sf{OSp}}

\begin{document}

\title{%
Supergravity as a constrained BF theory }
\author{R. Durka}
\email{rdurka@ift.uni.wroc.pl}\affiliation{Institute for Theoretical Physics,
University of Wroc\l{}aw, Pl.\ Maxa Borna 9, Pl--50-204 Wroc\l{}aw, Poland}
\author{J. Kowalski-Glikman}
\email{jkowalskiglikman@ift.uni.wroc.pl}\affiliation{Institute for Theoretical Physics,
University of Wroc\l{}aw, Pl.\ Maxa Borna 9, Pl--50-204 Wroc\l{}aw, Poland}
\author{M. Szczachor}
\email{misza@ift.uni.wroc.pl} \affiliation{Institute for Theoretical Physics,
University of Wroc\l{}aw, Pl.\ Maxa Borna 9, Pl--50-204 Wroc\l{}aw, Poland}

\date{\today}
\begin{abstract}
In this paper we formulate ${\cal N}=1$ supergravity as a constrained $BF$ theory with $OSp(4|1)$ gauge superalgebra. We derive the modified supergravity Lagrangian that, apart from the standard supergravity with negative cosmological constant, contains terms proportional to the (inverse of) Immirzi parameter. Although these terms do not change classical field equations, they might be relevant in quantum theory. We briefly discuss the perturbation theory around the supersymmetric topological vacuum.

\end{abstract}

\maketitle

\section{Introduction}
It is well known for quite some time that  gravity can be formulated as constrained topological field theory \cite{Smolin:1998qp}, \cite{Starodubtsev:2003xq}, \cite{Smolin:2003qu}, \cite{Freidel:2005ak}. The Lagrangian for such a theory contains two groups of terms. First we have the terms that describe a topological field theory of $BF$ type for the gauge group, which for gravity is chosen to be de Sitter $SO(4,1)$ or anti de Sitter $SO(3,2)$ group. These terms generate the topological vacuum of the theory. The remaining terms are responsible for the dynamics of gravity and are chosen in such a way so as to break the topological theory gauge symmetry down to the local $SO(d-1,1)$ Lorentz symmetry of gravity.

There are many advantages of such formulation of gravity.  First, as stressed in \cite{Freidel:2005ak}, it makes the kinetic term of the Lagrangian quadratic in fields, which makes the standard methods of quantum field theory applicable, contrary to the case of the Palatini formalism, in which the kinetic term is trilinear. Second, it opens an exciting possibility of a manifestly diffeomorphism invariance perturbative approach to quantum gravity (with and without matter sources) \cite{Freidel:2005ak}, \cite{KowalskiGlikman:2006mu}, in which the gauge breaking term is regarded as a perturbation around topological vacuum described by $BF$ theory. Third, this approach introduces the Immirzi parameter $\gamma$ \cite{Immirzi:1996di} to the theory in a natural way.

The presence of this parameter in the gravity Lagrangian was for many years overlooked because the corresponding term
\begin{equation}\label{0.0}
    \frac2{G\gamma}\, \epsilon^{\mu\nu\rho\sigma}R_{\mu\nu}{}^{ij} e_{\rho\, i}e_{\sigma\, j}\, ,
\end{equation}
 due to the second Bianchi identity, does not contribute to field equations when torsion vanishes. In quantum theory, however, $\gamma$ might be relevant, because it controls the rate of quantum fluctuations of torsion. Moreover, in spite of the fact that Immirzi parameter is not visible in field equations, its presence leads to modifications of the phase space structure of the theory, which in turn make it reappear in the spectra of Loop Quantum Gravity area and volume operators (see e.g., \cite{Ashtekar:2004eh}, \cite{Rovelli:2004tv}, \cite{Thiemann:2007zz}) and in the calculation of black hole entropy \cite{Rovelli:1996dv}. Further physical effects of Immirzi parameter are discussed in \cite{Perez:2005pm}  and \cite{Freidel:2005sn}.

In this paper we extend the construction of gravity as a constrained $BF$ theory to the case of ${\cal N}=1$ supergravity, generalizing the results reported in \cite{Ling:2000ss} to the case of the presence of Immirzi parameter. It turns out that in this case the term (\ref{0.0}) is replaced by its supersymmetrized counterpart, but it still does not influence classical equations of motion. However it might become relevant in quantum supergravity, and perhaps even in superstrings theory, of which the former is an low energy field-theoretical approximation.

The plan of this paper is as follows. In the next section we recall the construction of gravity as a constrained $BF$ theory. Next, in Section III and IV, we present the corresponding construction of ${\cal N}=1$ gravity, as a constrained $BF$ theory based on the $OSp(4|1)$ gauge superalgebra. Section V is devoted to the prove of supersymmetry invariance of the so obtained Lagrangian and to further discussion.  In the Appendix we collect some relevant definitions and formulas.

\section{Gravity as a constrained topological field theory}

 The construction of gravity as a constrained topological theory has its roots in the well known procedure of MacDowell and Mansouri \cite{MacDowell:1977jt} and has been developed recently by Smolin, Freidel, and Starodubtsev \cite{Smolin:1998qp}, \cite{Starodubtsev:2003xq}, \cite{Smolin:2003qu}, \cite{Freidel:2005ak}, \cite{Wise:2006sm}.

Let us recall briefly how this construction works, in the case of gravity with negative cosmological constant, which will be relevant for us later. In this case the gauge algebra is the $\SO(3,2)$ anti de Sitter algebra, with generators $M_{IJ}$ (our conventions can be found in the Appendix.) This algebra splits into its Lorentz and translational parts, generated by $M_{ij}$ and $P_i=M_{i5}$, respectively, and accordingly we can split the gauge field
\begin{equation}\label{0.1}
\mathbb{A_{\mu}}=\frac{1}{2}A_\mu{}^{ij} M_{ij}+A_\mu{}^{i5} M_{i5} =\frac{1}{2}\omega_\mu{}^{ij} M_{ij}+\frac{1}{\ell}e_\mu{}^{i} P_{i}
\end{equation}
with $\omega^{ij}$ being the Lorentz connection and $e_\mu^{i}$ identified with the tetrad. Notice that for dimensional reason, because the connection $\mathbb{A_{\mu}}$ has the canonical dimension $-1$ while the tetrad is dimensionless we have to introduce the parameter $\ell$ of dimension of length. As it will turn out this parameter is related to the cosmological constant.

Knowing the gauge field $\mathbb{A_{\mu}}$ we can built the curvature
\begin{equation}\label{0.2}
 \mathbb{F}_{\mu\nu}=\partial_\mu\mathbb{A}_{\nu}-\partial_\nu\mathbb{A}_{\mu}-i[\mathbb{A}_{\mu},\mathbb{A}_{\nu}]\, ,
\end{equation}
which, with the help of (\ref{a1}), can be again decomposed into Lorentz part
\begin{equation}\label{0.3}
    F_{\mu\nu}{}^{ij}= R_{\mu\nu}{}^{ij}- \frac1{\ell^2}\left(e_\mu{}^i\, e_\nu{}^j-e_\nu{}^i\, e_\mu{}^j \right)\, ,
\end{equation}
where
\begin{equation}\label{0.4}
    R_{\mu\nu}{}^{ij}=\partial_\mu\omega_\nu{}^{ij} - \partial_\nu\omega_\mu{}^{ij} +\omega_\mu{}^i{}_m\, \omega_\nu{}^{mj}-\omega_\nu{}^i{}_m\, \omega_\mu{}^{mj}
\end{equation}
is the Riemann tensor of the Lorentz connection $\omega_\mu{}^{ij}$, and the translational  part
$$
    F_{\mu\nu}{}^{i} =\frac1{\ell}\left( \partial_\mu e_\nu{}^{i}  + \omega_\mu{}^i{}_m\, e_\nu{}^m -\partial_\nu e_\mu^{i}  + \omega_\nu{}^i{}_m\, e_\mu{}^m\right) $$\begin{equation}\label{0.5}=\frac1{\ell}\left( D_\mu e_\nu{}^{i}-D_\nu e_\mu{}^{i}  \right)=\frac1{\ell}\, T_{\mu\nu}{}^i \, ,
\end{equation}
which is thus proportional to the torsion tensor.

  To construct the action we define another two form field $\mathbb{B}_{\mu\nu}$, which gauge-transforms in exactly the same way the curvature $\mathbb{F}_{\mu\nu}$ does. This field makes it possible to write down the topological $\SO(3,2)$-invariant Lagrangian
\begin{equation}\label{0.6}
    L^{top}= \epsilon^{\mu\nu\rho\sigma} \left( B_{\mu\nu}{}^{IJ}\, F_{\rho\sigma}{}_{IJ} - \frac\beta2\, B_{\mu\nu}{}^{IJ}\, B_{\rho\sigma}{}_{IJ}\right)
\end{equation}
with $\beta$ being a dimensionless parameter, to be related later to a combination of Newton's and cosmological constants, and Immirzi parameter. One checks (see \cite{Freidel:2005ak}) that the theory described by (\ref{0.6}) is indeed topological i.e.\ it does not contain any local degrees of freedom\footnote{To see this one notices that the theory described by (\ref{0.6}) is invariant not only with respect to the $\SO(3,2)$ gauge transformations but also with respect to translations of the field $\mathbb{B}_{\mu\nu}$, with the number of parameters just sufficient to make this field zero. Since it follows from the field equations that $\mathbb{F}_{\mu\nu}$ is proportional to $\mathbb{B}_{\mu\nu}$, the curvature vanishes, and the connection is flat. See \cite{Freidel:2005ak} for details.}. Its only solution is the anti de Sitter space and spaces that can be obtained from it by gauge transformations (for discussion see also \cite{KowalskiGlikman:2006mu}). As explained in \cite{Freidel:2005ak} in order to make this theory dynamical we have to add a term that breaks the symmetry down to the Lorentz subalgebra $\SO(3,1)$ of the original $\SO(3,2)$. This can be done by adding to the Lagrangian (\ref{0.6}) a gauge breaking term of the form
\begin{equation}\label{0.7}
    L^{gb}=-\frac\alpha4\, \epsilon^{\mu\nu\rho\sigma} \, \epsilon_{IJKL5} \, B_{\mu\nu}{}^{IJ}\, B_{\rho\sigma}{}^{KL}\, .
\end{equation}
with $\epsilon^{IJKLM}$ being the invariant epsilon symbol of $\SO(3,2)$. This term can be rewritten as
\begin{equation}\label{0.7a}
   L^{gb}=-\frac\alpha4\, \epsilon^{\mu\nu\rho\sigma} \, \epsilon_{ijkl} \, B_{\mu\nu}{}^{ij}\, B_{\rho\sigma}{}^{kl}\, ,
\end{equation}
from which it is clear that the Lagrangian  given by the sum of (\ref{0.6}) and (\ref{0.7a}) is indeed invariant under local Lorentz transformations, under which $F_{\mu\nu}{}^{ij}$ and $B_{\mu\nu}{}^{ij}$ transform as (anti-symmetric) tensors while $F_{\mu\nu}{}^{i5}\equiv F_{\mu\nu}{}^{i}$ and $B_{\mu\nu}{}^{i5}$ transform as vectors.

To see that the Lagrangian $L=L^{top}+L^{gb}$ is equivalent to the Lagrangian of general relativity we solve it for $\mathbb{B}_{\mu\nu}$ and substitute the result back to the Lagrangian. One finds
\begin{equation}\label{0.8}
    B_{\mu\nu}{}^{ij}=\frac{1}{\alpha^2+\beta^2}\left(\beta F_{\mu\nu}{}^{ij}-\frac{\alpha}2\, \epsilon^{ijkl}\, F_{\mu\nu}{}_{kl}\right)
\end{equation}
and
\begin{equation}\label{0.9}
    B_{\mu\nu}{}^{i5}=\frac1\beta\, F_{\mu\nu}{}^{i}\, .
\end{equation}
Substituting these equations into (\ref{0.6}) and (\ref{0.7}) we find that the resulting Lagrangian contains a combination of topological terms (corresponding to Euler, Pontryagin, and Nieh--Yan class) and the dynamical ones (we will derive this result while discussing the analogous supergravity construction below)
$$
L^{grav} = \frac{1}{G}\, \epsilon^{\mu\nu\rho\sigma}\epsilon_{ijkl} \left(R_{\mu\nu}{}^{ij}\, e_\rho{}^k\, e_\sigma{}^l - \frac\Lambda3\, e_\mu{}^{i}\,e_\nu{}^{j}\, e_\rho{}^k\, e_\sigma{}^l \right)
$$
\begin{equation}\label{0.10}
 +\frac{2}{\gamma G}   \epsilon^{\mu\nu\rho\sigma}\, R_{\mu\nu}{}^{ij}\, e_\rho{}_i\, e_\sigma{}_j \, ,
\end{equation}
where the Newton's constant $G$, cosmological constant $\Lambda$, and the Immirzi parameter $\gamma$ are related to the original parameters $\ell$, $\alpha$, and $\beta$ as follows
\begin{equation}\label{0.11}
    G = \frac{\alpha^2+\beta^2}{\alpha}\, \ell^2\, , \quad \Lambda = \frac3{\ell^2}\, , \quad \gamma = \frac{\beta}{\alpha}\, .
\end{equation}
The first two terms in the Lagrangian form nothing but the standard Einstein--Cartan Lagrangian. By virtue of the second Bianchi identity the last term does not contribute to the equations of motion on shell, when torsion vanishes, but it should be stressed that it is not topological and it influences the canonical structure of the theory.

\section{Gauging the super-algebra}

In this section we construct the building blocks of the constrained topological field theory of ${\mathcal N}=1$ supergravity.

As before, we associate with each generator of the $\OSp(4|1)$ algebra (whose defining relations can be found in the Appendix) a gauge field and a gauge transformation parameter. Since the canonical dimension of the gauge field is $-1$ and because the dimension of gravitino $\psi_\mu$ is $-3/2$, as in the bosonic case, we introduce the constant $\kappa$ of dimension $1/2$ to make the dimensions right. The gauge field is therefore
\begin{equation}\label{1.1}
\mathbb{A}_{\mu}=\frac{1}{2}\omega_\mu{}^{ij} M_{ij}+\frac{1}{\ell}e_\mu{}^{i} P_{i}+\kappa\bar{\psi}_\mu\, Q\, ,
\end{equation}
while the gauge transformation parameter reads
\begin{equation}\label{1.2}
    \Theta=\frac{1}{2}\Lambda^{ij}M_{ij}+\xi^i P_i+\bar{\epsilon}\, Q\, .
\end{equation}

The infinitesimal gauge transformations of the gauge field are defined in terms of the covariant derivative
\begin{equation}\label{1.3}
   \delta_\Theta \mathbb{A}_{\mu}\equiv D^{\mathbb{A}}_\mu \Theta=\partial_\mu \Theta -i[\mathbb{A}_{\mu}, \Theta]\, .
\end{equation}
Using this formula one can straightforwardly derive the supersymmetry transformations, to wit
\begin{equation}\label{1.4}
    \delta e_\mu{}^i = i\kappa\ell\, \bar\epsilon \gamma^i\psi_\mu\, , \quad \delta \omega_\mu{}^{ij} = -\kappa\,\bar\epsilon \gamma^{ij}\psi_\mu\, ,
\end{equation}
and
\begin{equation}\label{1.5}
    \delta \bar\psi_\mu{} =\frac1\kappa\, D_\mu\bar\epsilon= \frac1\kappa\, D^\omega_\mu\bar\epsilon + \frac{i}{2\kappa\ell}\, \,e_\mu{}^i\,\bar\epsilon\gamma_i    \, ,
\end{equation}
where
\begin{equation}\label{1.6}
    D^\omega_\mu\bar\epsilon=\partial_\mu\bar\epsilon-\frac14\omega_\mu{}^{ij}\, \bar\epsilon\gamma_{ij}
\end{equation}
is the covariant derivative of Lorentz connection $\omega$.

We define the curvature $\mathbb{F}_{\mu\nu}$ of the connection $\mathbb{A}_{\mu}$ as above
\begin{equation}\label{1.7}
    \mathbb{F}_{\mu\nu}=\partial_\mu\mathbb{A}_{\nu}-\partial_\nu\mathbb{A}_{\mu}-i[\mathbb{A}_{\mu},\mathbb{A}_{\nu}]\, .
\end{equation}
The curvature splits into bosonic and fermionic parts
\begin{equation}\label{1.8}
   \mathbb{F}_{\mu\nu}=\frac12\, F^{(s)}_{\mu\nu}{}^{IJ}\, M_{IJ} +  \bar{\mathcal F}_{\mu\nu} Q\, .
\end{equation}
Explicitly, we have
\begin{equation}\label{1.8a}
    F^{(s)}_{\mu\nu}{}^{ij} =F_{\mu\nu}{}^{ij}+ \kappa^2\, \bar\psi_\mu\gamma^{ij}\psi_\nu\, ,
\end{equation}
\begin{equation}\label{1.8b}
    F^{(s)}_{\mu\nu}{}^{i} =F_{\mu\nu}{}^{i} -i \kappa^2\, \bar\psi_\mu\gamma^{i}\psi_\nu\, ,
\end{equation}
where $F_{\mu\nu}{}^{ij}$ and $F_{\mu\nu}{}^{i}$ are bosonic curvatures given by (\ref{0.3}) and (\ref{0.5}) respectively. The fermionic curvature has the form
$$
  {\mathcal F}_{\mu\nu} =\kappa D_\mu\psi_\nu -\kappa D_\nu\psi_\mu$$ \begin{equation}\label{1.8c}
  =\kappa \left(D^\omega_\mu\psi_\nu - D^\omega_\nu\psi_\mu -\frac{i}{2\ell}\left(e_\mu{}^i\, \gamma_i\psi_\nu -e_\nu{}^i\, \gamma_i\psi_\mu\right)\right)\, ,
\end{equation}
where $D^\omega_\mu$ is the Lorentz covariant derivative, (see Appendix).

In what follows we will need the transformation rules for curvatures, which can be easily obtained from the following identities
$$
\delta \mathbb{F}_{\mu\nu} = D_\mu \delta\mathbb{A}_{\nu} -D_\nu \delta\mathbb{A}_{\mu} = [D_\mu, D_\nu]\Theta=-i[\mathbb{F}_{\mu\nu},\Theta]\,
.
$$
For the supersymmetry transformation we have therefore
\begin{equation}\label{1.9}
    \delta_\epsilon F^{(s)}_{\mu\nu}{}^{i} = i\bar\epsilon \gamma^{i} {\mathcal F}_{\mu\nu}, \quad  \delta_\epsilon F^{(s)}_{\mu\nu}{}^{ij} = -\bar\epsilon \gamma^{ij} {\mathcal F}_{\mu\nu}\, ,
\end{equation}
\begin{equation}\label{1.10}
    \delta_\epsilon \bar{\mathcal F}_{\mu\nu} = -\frac14\, F^{(s)}_{\mu\nu}{}^{ij}\, \bar\epsilon\gamma_{ij}+\frac{i}{2}\, \,F^{(s)}_{\mu\nu}{}^i\,\bar\epsilon\gamma_i\, .
\end{equation}

With these technical tools at hands we can now address the problem of constructing of the supersymmetric extension of the Lagrangian (\ref{0.6}) and (\ref{0.7}). Let us first consider the topological theory, whose bosonic part is given by (\ref{0.6}). We introduce the fermionic partner of the field $B_{\mu\nu}{}^{IJ}$ which we denote as ${\mathcal B}_{\mu\nu}$ so that the Lagrangian reads
$$
L^{(sugra-top)} =  L^{(sugra-top,b)}- L^{(sugra-top,f)}
$$$$
=\epsilon^{\mu\nu\rho\sigma} \left( B_{\mu\nu}{}^{IJ}\, F^{(s)}_{\rho\sigma}{}_{IJ} - \frac\beta2\, B_{\mu\nu}{}^{IJ}\, B_{\rho\sigma}{}_{IJ}\right)
$$
\begin{equation}\label{1.11}
   -4\,\epsilon^{\mu\nu\rho\sigma} \left( \bar{\mathcal B}_{\mu\nu}{\mathcal F}_{\rho\sigma}- \frac\beta2\,\bar{\mathcal B}_{\mu\nu}{\mathcal B}_{\rho\sigma}\right)\, .
\end{equation}
This Lagrangian is invariant under local supersymmetry if the components of the field ${\mathbb B}=(B, {\mathcal B})$ transform as follows
\begin{equation}\label{1.12}
\delta_\epsilon B_{\mu\nu}{}^{i} = i \bar\epsilon \gamma^{i} {\mathcal B}_{\mu\nu},\quad \delta_\epsilon B_{\mu\nu}{}^{ij} = -\bar\epsilon \gamma^{ij} {\mathcal B}_{\mu\nu},
\end{equation}
\begin{equation}\label{1.12a}
    \delta_\epsilon \bar{\mathcal B}_{\mu\nu} =-\frac14\,  B_{\mu\nu}{}^{ij}\, \bar\epsilon \gamma_{ij} + \frac{i}{2}\,B_{\mu\nu}{}^{i}\, \bar\epsilon \gamma_{i}\, .
\end{equation}

The gauge breaking term (\ref{0.7a}) is invariant only under the action of the $\SO(3,1)$ Lorentz subalgebra of the original gauge algebra $\SO(3,2)$. Its supersymmetric extension is expected to be
\begin{equation}\label{1.13}
 L^{sugra-gb} =   -\frac\alpha4\, \epsilon^{\mu\nu\rho\sigma} \left( \epsilon_{ijkl} \, B_{\mu\nu}{}^{ij}\, B_{\rho\sigma}{}^{kl} -8i \, \bar{\mathcal B}_{\mu\nu}\gamma^5{\mathcal B}_{\rho\sigma}\right)\, .
\end{equation}
This term, however, is not invariant under the supersymmetry transformations (\ref{1.12}), (\ref{1.12a}) since under the latter the second term in (\ref{1.13}) gets the contribution of the form
\begin{equation}\label{1.14}
-2\alpha\epsilon^{\mu\nu\rho\sigma} B_{\mu\nu}{}^{i}\bar\epsilon\gamma_i\gamma^5{\mathcal B}_{\rho\sigma}
\end{equation}
that does not cancel with the supersymmetry  transformation of the first term. As we discuss below this  breaking of supersymmetry, related to the breaking of de Sitter group down to its Lorentz subgroup, does not prevent the final action from having the local supersymmetry invariance.

\section{The supergravity Lagrangian}

Let us now check explicitly that our procedure indeed provides the lagrangian of ${\mathcal N}=1$ supergravity. Our starting point will be the sum of the terms (\ref{1.11}) and (\ref{1.14}).
The equations of bosonic $B_{\mu\nu}{}^{IJ}$ result in expressions analogous to (\ref{0.8}) and (\ref{0.9})
$$
    B_{\rho\sigma}{}^{ij}=\frac{\beta}{\alpha^2+\beta^2} \left(F_{\rho\sigma}{}^{ij}(A)+\kappa^2\bar{\psi}_\rho\,\gamma^{ij}\,\psi_\sigma\right)
    $$
    \begin{equation}\label{2.1}
    -\frac{\alpha}{2(\alpha^2+\beta^2)}\, \left(F_{\rho\sigma}{}^ {kl}(A)+\kappa^2\bar{\psi}_\rho\,\gamma^{kl}\,\psi_\sigma\right)\epsilon^{ij}{}_{kl}\, ,
\end{equation}
\begin{equation}\label{2.2}
   B_{\rho\sigma}{}^{i}=\frac{1}{\beta}\left(F_{\rho\sigma}{}^{i}(A)-i\kappa^2\bar{\psi}_\rho\,\gamma^{i}\,\psi_\sigma\right)\, ,
\end{equation}
 while for their fermionic counterpart we obtain
 \begin{equation}\label{2.3}
    \mathcal{B}=\frac{1}{\alpha^2+\beta^2}\left(\beta \mathcal{F}-i\alpha\,\gamma_5\, \mathcal{F}\right)\, .
 \end{equation}
 Substituting these expressions back to the Lagrangian
$$
L^{F}=\frac{1}{(\alpha^2+\beta^2)}\int\epsilon^{\mu\nu\rho\sigma}(\frac{\beta}{2}\bar{\mathcal{F}}_{\mu\nu}
\,\mathcal{F}_{\rho\sigma}-\frac{i\alpha}{2}\bar{\mathcal{F}}_{\mu\nu} \,\gamma_5\,\mathcal{F}_{\rho\sigma})
$$
 and disregarding terms that vanish due to the identities
 $$
 \epsilon^{\mu\nu\rho\sigma}\, \bar\psi_\mu\, \Gamma\, \psi_\nu\, \Gamma\, \psi_\rho =0\, ,
 $$
where $\Gamma$ is an arbitrary combination of $\gamma$ matrices,
$$
\epsilon^{\mu\nu\rho\sigma}\,\bar\psi_\mu\, \Gamma^A\, \psi_\nu=0, \quad \Gamma^A=(1,\gamma^5,\gamma^5\gamma^i)\, ,
$$
and expressions (\ref{1.8a}), (\ref{1.8b}), (\ref{1.8c}) we find the Lagrangian that can be decomposed into two types of terms:

 \begin{itemize}
\item Bosonic
$$ L^{(B)}=   \left(\frac{1}{\beta\ell^2}T_{\mu\nu\,i}T^{i}_{\rho\sigma}+\frac{\beta}{2(\alpha^2+\beta^2)}F_{\mu\nu\,ij}F_{\rho\sigma}{}^{ij}\right)\epsilon^{\mu\nu\rho\sigma}
$$ \begin{equation}\label{2.4}   -\frac{\alpha}{4(\alpha^2+\beta^2)}F_{\mu\nu\,ij}F_{\rho\sigma\,kl}\epsilon^{ijkl}\epsilon^{\mu\nu\rho\sigma}\, ;
\end{equation}
\item Fermionic
 $$
L^{(F)}= -\frac{8\alpha\kappa^2}{(\alpha^2+\beta^2)\ell} \bar{\psi}_\mu\,\gamma_5\,\gamma_{i}\,e^{i}_{\nu}D^\omega_{\rho}\psi_\sigma\, \epsilon^{\mu\nu\rho\sigma}
$$
\begin{equation}\label{2.5}
+    \frac{4i\alpha\kappa^2}{(\alpha^2+\beta^2)\ell^2} \bar{\psi}_\mu\,\gamma_5\,\gamma_{ij}\,e^{i}_{\nu}e^{j}_{\rho}\,\psi_\sigma \, \epsilon^{\mu\nu\rho\sigma}
\end{equation}
$$
-\frac{2i\kappa^2\alpha^2}{\ell\beta(\alpha^2+\beta^2)}\, \bar{\psi}_\mu\, \gamma_{i}\, \psi_\nu T_{\rho\sigma}^{i}\,\epsilon^{\mu\nu\rho\sigma}\, .
$$
 \end{itemize}
Using the expansion of the curvature $F_{\rho\sigma}{}^{ij}$ (\ref{0.3}) one sees that the first two terms in (\ref{2.4}) provide a combination of Pontryagin and  Nieh-Yan \cite{Nieh:1981ww} (see also  \cite{Chandia:1997hu})
  \begin{equation}\label{2.6}
   N= \epsilon^{\mu\nu\rho\sigma}\left(T_{\mu\nu}{}^i\, T_{\rho\sigma}{}_i- 2 R_{\mu\nu}{}^{ij}\, e_\rho{}_i\, e_\sigma{}_j\right)
\end{equation}
classes along with the term proportional to $\epsilon^{\mu\nu\rho\sigma}\, R_{\mu\nu}{}^{ij}\, e_\rho{}_i\, e_\sigma{}_j$; the remaining one reproduces the Euler class and the Einstein--Cartan Lagrangian (\ref{0.10}) and the relations (\ref{0.11}).

Let us now turn to the fermionic terms. We fix $\kappa$ so as to make the coefficient of the gravitino kinetic term equal $-1/2$ and thus
\begin{equation}\label{2.7}
    \kappa^2=\frac1{16}\, G\, \sqrt{\frac\Lambda3}=\frac{G}{16\ell}\, .
\end{equation}
Thus our Lagrangian is the sum of standard Lagrangian $L^{sugra}$ of supergravity with cosmological constant \cite{Townsend:1977qa}\footnote{The relation between our parameters and that of Townsend is $\kappa_{Townsend} =\sqrt{G}/4$, $\lambda =4\sqrt{\Lambda/3G}$. Also in \cite{Townsend:1977qa} the supersymmetry transformation parameter is not dimensionless, but instead $\epsilon_{Townsend} = \sqrt\ell \epsilon_{our}$.}
given by
$$
L^{sugra}=\frac{1}{G}\, \epsilon^{\mu\nu\rho\sigma}\epsilon_{ijkl} \left(R_{\mu\nu}{}^{ij}\, e_\rho{}^k\, e_\sigma{}^l - \frac\Lambda3\, e_\mu{}^{i}\,e_\nu{}^{j}\, e_\rho{}^k\, e_\sigma{}^l \right)
$$
\begin{equation}\label{2.8}
    -\epsilon^{\mu\nu\rho\sigma} \left(\frac{1}{2} \, \bar{\psi}_\mu\,\gamma_5\,\gamma_{i}\,e^{i}_{\nu}D^\omega_{\rho}\psi_\sigma\, -\frac{i}{4\ell} \, \bar{\psi}_\mu\,\gamma_5\,\gamma_{ij}\,e^{i}_{\nu}e^{j}_{\rho}\,\psi_\sigma \right)
\end{equation}
and the  term $L^{add}$, which is the supersymmetric counterpart of the term in the second line of (\ref{0.10}), to wit
\begin{equation}\label{2.9}
L^{add}=       \epsilon^{\mu\nu\rho\sigma}\left(\frac{2}{\gamma G} \, R_{\mu\nu}{}^{ij}\, e_\rho{}_i\, e_\sigma{}_j - \frac{i}{4\gamma}\,\bar{\psi}_\mu\, \gamma_{i}\, \psi_\nu \, D^\omega_\rho e_\sigma{}^i\right)\, .
\end{equation}

This is the main result of our paper. In the next section we will discuss some properties of this Lagrangian.

\section{Discussion}

Our first task after deriving the form of the Lagrangian will be to check that the action obtained from (\ref{2.8}), (\ref{2.9}) is indeed invariant under supersymmetry. To do that we will make use of the 1.5 formalism (see \cite{VanNieuwenhuizen:1981ae} and references therein), which combines the virtues of the first ($\omega$ is an independent field) and second ($\omega=\omega(e,\psi)$) order formalisms. The idea is as follows. Our action $I$, being the integral of the Lagrangian can be thought of as a functional $I(e,\psi; \omega(e,\psi))$ and its variation is
$$
    \delta I = \delta e\, \left.\frac{\delta I}{\delta e}\right|_{\psi, \omega(e,\psi)} + \delta \psi\, \left.\frac{\delta I}{\delta \psi}\right|_{e, \omega(e,\psi)}$$
\begin{equation}+\left.\frac{\delta I}{\delta \omega}\right|_{e,\psi}\left(\delta e\,\frac{\delta \omega(e,\psi)}{\delta e}+ \delta \psi\,\frac{\delta \omega(e,\psi)}{\delta \psi}\right) \, .\label{3.1}\end{equation}
But if $\omega$ satisfies its own field equations the last term in (\ref{3.1}) vanishes identically, because $\delta I/\delta\omega=0$ for $\omega$ satisfying its own field equation. In other words we need to vary only the gravitino and tetrad fields, taking into account, where necessary, the conditions coming from the Lorentz connection field equations.

The first step therefore in checking the supersymmetry is to find the form of the $\omega$ field equations. They read
\begin{equation}\label{3.2}
    \left(\epsilon^{ijkl}\, F^{(s)}{}_{\nu\rho\, k}\, e_{\sigma\, l} + \frac1\gamma\, F^{(s)}{}_{\nu\rho}{}^{[i}\, e_{\sigma}{}^{j]}\right)\epsilon^{\mu\nu\rho\sigma}=0\, ,
\end{equation}
where the `supertorsion' $F^{(s)}$ is defined by
\begin{equation}\label{3.3}
    F^{(s)}{}_{\nu\rho}{}^{i}=F_{\nu\rho}{}^{i}-i\kappa^2\, \bar\psi_\nu\gamma^i\psi_\rho\, .
\end{equation}
It follows from (\ref{3.2}) that the supertorsion vanishes\footnote{To see this one has to multiply (\ref{3.2}) by $\epsilon_{ijmn}$ and use the fact that tetrad is invertible.} unless $\gamma^2=-1$. Form this one can find the Lorentz connection as usual (see \cite{VanNieuwenhuizen:1981ae}).

Since the Lorentz connection field equations are the same as in the standard case of ${\cal N}=1$ supergravity, it is just a matter of repeating verbatim the steps described in \cite{VanNieuwenhuizen:1981ae}, \cite{Townsend:1977qa} to see that (\ref{2.8}) is indeed supersymmetric (up to the usual total derivative term). What remains therefore is to check that $L^{add}$ (\ref{2.9}) is supersymmetric as well. But this is also quite straightforward.

In fact we will prove a much stronger result namely that {\em if supertorsion is zero, for arbitrary $\delta e$, $\delta \psi$ the variation of $L^{add}$ vanishes}. This not only proves the supersymmetry invariance but also shows that the Lagrangian $L^{add}$ does not contribute to the field equations. Consider the second term in (\ref{2.9}) first. Varying  gravitino we find
$$
 - \frac{i}{2\gamma}\,\epsilon^{\mu\nu\rho\sigma}\,  \delta\bar{\psi}_\mu\, \gamma_{i}\, \psi_\nu \, D^\omega_\rho e_\sigma{}^i\, .
$$
But since $D^\omega{}_{[\rho} e_{\sigma]}{}^i\sim \bar\psi_\rho\gamma^i\psi_\sigma$ from vanishing of supertorsion (\ref{3.3}), using the identity
$$
\epsilon^{\mu\nu\rho\sigma}\, \gamma_i\psi_\nu\,\bar\psi_\rho\gamma^i\psi_\sigma=0
$$
we see that this expression vanishes. Thus it remains to check the variation of tetrad. We find
\begin{equation}\label{3.4}
    \epsilon^{\mu\nu\rho\sigma}\left(\frac{4}{\gamma G} \, R_{\mu\nu}{}^{ij}\, e_\rho{}_i\,  \delta e_\sigma{}_j - \frac{i}{4\gamma}\,\bar{\psi}_\mu\, \gamma_{i}\, \psi_\nu \, D^\omega_\rho\delta e_\sigma{}^i\right)\, .
\end{equation}
To see that this expression is indeed zero we first make use of the second Bianchi identity
$$
\epsilon^{\mu\nu\rho\sigma}\, R_{\mu\nu}{}^{ij}\, e_\rho{}_i=-2\epsilon^{\mu\nu\rho\sigma}\, D^\omega_\mu\, D^\omega_\nu e_\rho{}^j\, .
$$
Therefore (\ref{3.4}) can be (up to the total derivative) rewritten as
$$
\epsilon^{\mu\nu\rho\sigma}\left(\frac{8}{\gamma G} \, D^\omega_\nu e_\rho{}^j\, D^\omega_\mu \delta e_\sigma{}_j - \frac{i}{4\gamma}\,\bar{\psi}_\mu\, \gamma_{i}\, \psi_\nu \, D^\omega_\rho\delta e_\sigma{}^i\right)\, .
$$
which, with the help of the supertorsion equation and (\ref{2.7}) can be easily seen to vanish. This completes not only the prove of supersymmetry, but also shows that the terms (\ref{2.9}) do not contribute to field equations. However it should be stressed that although invisible classically the term $L^{add}$ might be relevant in quantum theory like QCD theta term.

It is worth noticing that the proof of supersymmetry of the final supergravity Lagrangian $L^{sugra} + L^{add}$ makes it possible to resolve the puzzle that we encounter at the end of Section III. Namely if we make use of the fact that $B_{\mu\nu}{}^{i}$ equals supertorsion and that the latter vanishes we see that  the expression (\ref{1.14}) is zero. This is why the apparent lack of supersymmetry of the constrained theory does not prevent the final one from being supersymmetric.

Finally let us turn to the brief discussion of perturbation theory around (super-) topological vacuum. To this end let us return to the original constrained $BF$ theory, given by (\ref{1.11}), (\ref{1.13})  and consider the case $\alpha=0$. The bosonic field equations are
\begin{equation}\label{3.5}
    F_{\mu\nu}{}^{IJ} = \frac1\beta\, B_{\mu\nu}{}^{IJ}, \quad \epsilon^{\mu\nu\rho\sigma}\, D_\nu B_{\rho\sigma}{}^{IJ}=0\, ,
\end{equation}
where due to the Bianchi identity the first equation implies the second one. Notice now that the bosonic part of (\ref{1.11}) is invariant not only with respect to $SO(3,2)$ gauge symmetry but also has the `translational', topological symmetry generated by $SO(3,2)$ Lie algebra valued local gauge parameter $\varphi_{\mu}{}^{IJ}$ \cite{Freidel:2005ak}
\begin{equation}\label{3.6}
    \delta B_{\mu\nu}{}^{IJ}=D_{[\mu} \varphi_{\nu]}{}^{IJ}, \quad \delta A_{\mu}{}^{IJ}=\beta\, \varphi_{\mu}{}^{IJ}\, .
\end{equation}
It follows that the bosonic part of the Lagrangian (\ref{1.11}) is independent of $A_{\mu}{}^{IJ}$, which is the sign of topological invariance (see \cite{Cattaneo:1995tw} for details). By supersymmetry, the same holds for the full Lagrangian (\ref{1.11}). Thus one can device the perturbation theory around a supersymmetric topological vacuum exactly in the way described in \cite{KowalskiGlikman:2008fj} with the only difference being that now the path integral is over $SP(4|1)$ (super-) connections, instead of the $SO(3,2)$ (or $SO(4,1)$) ones. Making this idea precise will be a subject of our future research.

There are many interesting open problems which deserve investigation. It would be interesting to derive ${\cal N}=2$ and possibly also higher supergravity theories using the constrained $BF$ theory of $SP(4,N)$ superalgebra, and their coupling to (supersymmetric matter). It would be also of interest to understand the role of auxiliary fields in the $BF$ formalism. These problems are currently under investigations.


\begin{acknowledgments}
  JKG is supported in part by research projects N202 081 32/1844 and NN202318534 and Polish Ministry of Science and Higher Education grant 182/N-QGG/2008/0.
\end{acknowledgments}

\appendix*
\section{Useful formulas}

In this Appendix we present our conventions, borrowing them from  \cite{Nicolai:1984hb}, and we collect formulas that are used in the main text.

We start with the supersymmetry algebra $\OSp(1|4)$. Its bosonic part can be obtained from the $\SO(3,2)$ algebra
\begin{eqnarray}
[M_{IJ},M_{MN}] &=&  i(\eta_{IN}M_{JM}+ \eta_{JM}M_{IN} \nonumber\\
  &-&\eta_{IM}M_{JN}-\eta_{JN}M_{IM})\, ,\label{a1}
\end{eqnarray}
  where $\eta_{IJ}$ is the metric tensor of signature $(+,-,-,-,+)$, and the capital indices $I,J,\ldots$ take the values $i,j,\ldots = 0,\dots, 3$ and $4$. We also define the Levi-Civita epsilon symbol $\epsilon^{IJKLM}$ by $\epsilon^{01234}=1$ so that $\epsilon^{ijkl4}=\epsilon^{ijkl}$. Decomposing the $\SO(3,2)$ generators $M_{IJ}$ into Lorentz generators $M_{ij}$ and momenta $P_i =M_{i4}$ we find
\begin{equation}\label{a2}
[M_{ij},M_{mn}]=i(\eta_{in}M_{jm}+\eta_{jm}M_{in}-\eta_{im}M_{jn}-\eta_{jn}M_{im})\, ,
\end{equation}
\begin{equation}\label{a3}
[P_i,M_{mn}]=i(\eta_{im}P_{n}-\eta_{in}P_{m})\, ,
\end{equation}
\begin{equation}\label{a4}
[P_{i},P_{j}]=-i\eta_{44}M_{ij}=-iM_{ij}\, .
\end{equation}
The $\gamma$  matrices satisfy the standard Clifford algebra
\begin{equation}\label{a4a}
    \{\gamma^i, \gamma^j\}=2\eta^{ij}, \quad \eta^{ij}=\mbox{diag}(+,-,-,-)\, .
\end{equation}
Explicitly, we have
$$
\gamma^{0} = \left(
\begin{array}{cc}
1 & 0\\
0 & -1
\end{array}
\right)
\quad
\gamma^{a} = \left(
\begin{array}{cc}
0 & \sigma^a\\
-\sigma^a & 0
\end{array}
\right)\quad \mbox{for } a=1,2,3
$$
One checks that the following combination of $\gamma$ matrices
\begin{equation}\label{a4b}
    m_{i4}=-\frac{i}2\, \gamma_i, \quad m_{ij}=\frac12 \gamma_{ij}=\frac14\, [\gamma_i, \gamma_j]
\end{equation}
forms a representation of the $\SO(3,2)$ (or, more precisely, of the algebra of its covering group $\sf{Sp}(4)$) (\ref{a1}).

The supersymmetry generator $Q$ transforms as a (Majorana) spinor with respect to the $\SO(3,2)$
\begin{equation}\label{a5}
    [M_{IJ},Q_\alpha]=-i(m_{IJ})_{\alpha}^{~~\beta}\, Q_\beta \, ,
\end{equation}
i.e.\
\begin{equation}\label{a6}
   [M_{ij}, Q] =-\frac{i}2\, \gamma_{ij}\, Q, \quad [P_i, Q] = -\frac12\, \gamma_i\, Q
\end{equation}
Finally the anicommutator of two supersymmetry generators reads
\begin{equation}\label{a7}
    \{Q_\alpha,\bar Q_\beta\}=im^{IJ}_{\alpha\beta}\, M_{IJ}\, ,
\end{equation}
which can be split to
\begin{equation}\label{a8}
    \{Q_\alpha,\bar Q_\beta\}=\frac{i}{2}(\gamma^{ij})_{\alpha\beta}\, M_{ij} +  \gamma^i\, P_i\, .
\end{equation}

The $\OSp(1|4)$ algebra given by (\ref{a2})--(\ref{a4}) and (\ref{a6}), (\ref{a8}) can be contracted to the super-Poincar\'e algebra as follows. We rescale $P_i \rightarrow \ell\, P_i$ and $Q \rightarrow \sqrt \ell\, Q$ and then let $\ell\rightarrow\infty$. As a result we find that the commutators (\ref{a2}), (\ref{a3}) do not change; the right hand side of (\ref{a4}) and the second commutator in (\ref{a6}) vanish, while in (\ref{a8}) only the second term on the right hand side survives.

The covariant derivative is defined as follows
\begin{eqnarray}
    D_{\mu}\bar{\psi}_\nu&=& \partial_{\mu}\bar{\psi}_\nu-\frac{1}{4}\omega^{ij}_\mu\,\bar{\psi}_\nu\,\gamma_{ij}+\frac{i}{2\ell} e^{i}_{\mu}\,\bar{\psi}_\nu\,\gamma_{i}\nonumber\\
D_{\mu}\psi_\nu&=& \partial_{\mu}\psi_\nu+\frac{1}{4}\omega^{ij}_\mu\,\gamma_{ij}\,\psi_\nu-\frac{i}{2\ell} e^{i}_{\mu}\,\gamma_{i}\,\psi_\nu\, .
\end{eqnarray}

\end{document}